\documentclass[%
 reprint,
 amsmath,amssymb,
 aps,
 pra,
nobalancelastpage
]{revtex4-2}

\usepackage{graphicx}
\usepackage{dcolumn}
\usepackage{bm}
\usepackage{lipsum}
\usepackage{xcolor}
\usepackage{dsfont}
\usepackage{amsmath}
\usepackage{amssymb}
\usepackage[colorlinks=true, linkcolor=blue,citecolor=blue]{hyperref}

\newcommand\munu{{\mu\nu}}

\newcommand\cala{\mathcal A}

\newcommand\eye{\tensor 1}

\newcommand\uvec[1]{\hat {#1}}

\newsavebox{\mySaveBoxMath} 
\newsavebox{\mySaveBoxText} 

\renewcommand{\vec}[1]{\boldsymbol{#1}}
\renewcommand{\uvec}[1]{\hat{\boldsymbol{#1}}}
\renewcommand{\tensor}[1]{\underline{\boldsymbol{#1}}}
\newcommand\utensor[1]{\tensor{\hat {#1}}}

\newcommand{\compactcross}[1]{\hspace[-0.1em]\times\hspace[-0.1em]}
\renewcommand{\Im}{\operatorname{Im}}
\renewcommand{\Re}{\operatorname{Re}}
\newcommand{\env}{\vec{\mathcal E}}

\newcommand{\ham}{\mathcal H}

\begin{document}

\preprint{APS/123-QED}

\title{Spin-redirection Berry phase with planar rays}

\author{Aymeric Braud}
 \email{aymeric.braud@laplace.univ-tlse.fr}
\author{Renaud Gueroult}
 \affiliation{LAPLACE, Université de Toulouse, CNRS, INPT, UPS, 31062 Toulouse, France}

\date{\today}

\begin{abstract}
Geometric or Berry phases are fundamental manifestations that appear in many areas of physics. They arise from the geometry of the space describing the properties of multi-component wave fields. An important example for electromagnetic waves is the spin-redirection Berry phase associated with the evolution of the spin direction. Because this effect has traditionally been studied in isotropic media where the spin is aligned with the ray trajectory, it has become commonly assumed that this spin-redirection Berry phase requires nonplanar rays. Here we show that a spin-redirection phase can in fact arise along a planar ray if the spin evolves along the ray. We expose this effect through the singular example of a moving unmagnetized plasma, and demonstrate how this behavior can more generally arise from a finite transverse spin. In identifying this new spin-redirection mechanism our work not only provides the tools to discover additional manifestations of SOIs in nature, but also uncovers supplemental degrees of freedom to harness SOIs to control light.
\end{abstract}

\maketitle

\section{Introduction}

Electromagnetic waves carry both spin~\cite{Beth1936} and orbital angular momentum~\cite{Poynting1909,Allen1992}, associated, respectively, to the wave polarization and
spatial degrees of freedom. Because of the vectorial nature of Maxwell’s equations,
polarization is, however, not independent of the
orbital degrees of freedom: one influences the other and reciprocally ~\cite{liberman_spin-orbit_1992,Allen1996,bliokh_geometrodynamics_2009}. This coupling between
light's spin and orbital degrees of freedom is known as
spin-orbit interactions (SOIs) of light~\cite{bliokh_spinorbit_2015},  analogously to SOIs of quantum particles~\cite{Mathur1991} or of electrons in solids~\cite{Berard2006}.

When photons or electrons experience SOIs, they acquire an additional phase, known as geometric or Berry phase~\cite{berry_quantal_1984,berry_budden_1990}. This Berry phase, by adding to the dynamical phase, has been shown to play a key role in a remarkably wide range of physical contexts~\cite{cohen_geometric_2019}. Focusing on light, it notably captures the parallel transport of light’s polarization state~\cite{bliokh_spinorbit_2015}. A prominent example is the rotation of polarization known as Rytov's rotation~\cite{rytov_sur_1938,Vladimirskiy1941}, associated with
the spin-redirection or Rytov-Vladimirsky-Berry phase. This spin redirection effect notoriously occurs in response to variations of the wavevector direction, as extensively studied in inhomogeneous isotropic media~\cite{liberman_spin-orbit_1992,bliokh_geometrodynamics_2009,Torabi2012}, or in a helically wound optical fiber~\cite{Ross1984,Tomita1986}. Because in both of these archetypal examples the redirection is due to the nonplanarity of the wave trajectory (i.e. ray), it is commonly assumed that spin-redirection SOI and its manifestation on polarization require variations of the direction of propagation and notably nonplanar rays~\cite{bliokh_coriolis_2008,Liu2016,Bliokh2019,cisowski_colloquium_2022,Sheng2023}.


Fundamentally though, spin redirection simply requires that the polarization plane varies along the ray~\cite{cohen_geometric_2019}. For transverse waves the spin is classically parallel to the wavevector, which in isotropic media is itself parallel to the group velocity, so that spin redirection can only come from a redirection of the ray, and more particularly from nonplanar rays. On the other hand, it is now well established that waves can exhibit significant transverse spin~\cite{Bliokh2015a}. The spin can then locally deviate from the group velocity, suggesting the existence of an additional mechanism for spin redirection. Here we show that spin-redirection SOI can in fact arise even with planar (i.e. torsionless) rays, on the condition that the polarization plane evolves along the ray. We expose this effect through the striking example of a moving unmagnetized plasma, for which the spin is redirected by motion even though the ray remains straight, just like in a static homogeneous medium. Importantly, because this new contribution to the spin-redirection phase will in general add up to the ray-redirection contribution classically known for nonplanar rays, our results identify additional degrees of freedom to control SOIs of light. Because finite transverse spin waves~\cite{Bliokh2015a} have been found to be common to numerous systems including metamaterials~\cite{Peng2019}, this creates new opportunities for light manipulation~\cite{Cardano2015,Ling2017,jisha_geometric_2021}. Furthermore, SOIs have recently been suggested to play an important role in different processes, such as depolarization in the solar wind and the interstellar medium~\cite{Zhang2024}, or the dynamics of solar acoustic modes~\cite{Leclerc2025} and gravitational waves~\cite{Oancea2024}. In identifying a connection between this novel mechanism for spin-redirection SOI and finite transverse spin waves~\cite{Bliokh2015a}, as found for instance in magnetized plasmas, our findings suggest that SOIs may in fact be at play in a broader range of environments.


\begin{figure*}
    \centering
    \begin{picture}(0,0)
        \put(-0.5\linewidth,-7.545cm){\textbf{(a)}} 
    \end{picture}
    \begin{picture}(0,0)
        \put(0.029\linewidth,-7.545cm){\textbf{(b)}} 
    \end{picture}
    \includegraphics[width=\linewidth]{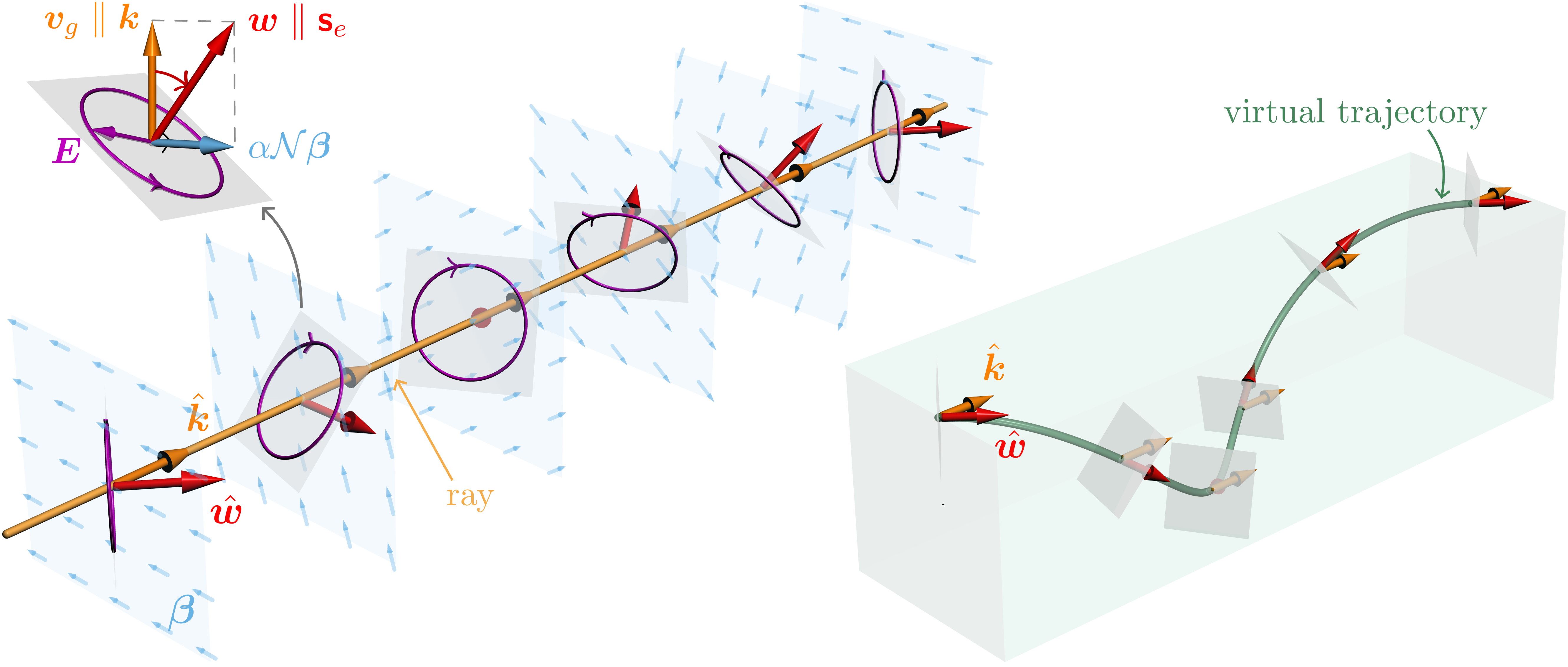}
    \caption{(a) Ray trajectory (orange) and polarization plane (grey plane) of a wave in a moving (blue arrows) unmagnetized plasma. The trajectory, which is not affected by motion, remains straight. On the other hand, since the vector $\uvec w$ (red) which is parallel to the electric spin density $\vec {\mathsf{s}}_e$ is redirected by motion, the polarization plane orientation varies along the ray. (b) The same polarization plane redirection is obtained when associating with this ray a nonplanar virtual trajectory (green) defined by the tangent vector $\uvec w$. }
    \label{fig:1}
\end{figure*}

\section{Results}
\subsection{Polarization plane redirection along straight rays}
Let us first show how the motion of an unmagnetized plasma affects the polarization plane of electromagnetic waves propagating through it, while leaving the ray trajectory unaltered and therefore straight. For that we consider a cold homogeneous plasma with a refractive index at rest $n'(\omega)=(1-\omega_p^2/\omega^2)^{1/2}$, with $\omega_p$ the plasma frequency, and write $\vec v(\vec x)=\vec \beta(\vec x)c$ the local plasma bulk velocity as seen in the laboratory frame with $c$ the speed of light.


In geometrical optics waves are described by rays~\cite{born_principles_2019,tracy_ray_2014}. To lowest order in eikonal expansion the wave polarization and trajectory are determined by the dispersion matrix $\tensor D$ obtained by assuming plane waves $e^{i(\omega t-\vec k\cdot\vec x)}$ in the equations of dynamics. Working with the electric field $\vec E$, we show using covariant formalism in Appendix~\ref{app:pola_plane_redir} that for a moving unmagnetized plasma
\begin{equation}
    \tensor D=\tensor M\ham_0+\vec w\vec w^T
    \label{Eq:D0_simplified}
\end{equation}
where $\ham_0(\vec k,\omega)={(\omega^2-\omega_p^2)}/{c^2}-k^2$ is the usual dispersion function for an unmagnetized cold plasma at rest~\cite{rax_physique_2005} and
\begin{equation}
    \vec w(\vec x,\vec k,\omega)=\vec k+\alpha\mathcal N(\omega)\vec\beta(\vec x),
    \label{Eq:w_simplified}
\end{equation}
is a vector that we find is parallel to the Poynting vector $\vec \Pi=\Re\left(\vec E\times\vec H^*\right)/2$. Here $\tensor M$ and $\alpha$ are, respectively, tensor and scalar functions of $(\vec k,\omega,\vec \beta)$, and $\mathcal N(\omega)=n'(\omega)-1/n'(\omega)$ is the Fresnel drag coefficient~\cite{fermi_rotation_1923}. Noting that $\vec w\vec w^T$ is a projector on $\vec w$, Eq.~\eqref{Eq:D0_simplified} shows that a moving unmagnetized plasma supports two degenerate modes in the plane normal to $\vec w$ which verify the dispersion relation $\ham_0=0$. Although singular, the fact that these modes satisfy the same dispersion relation as the propagative modes in this same plasma but at rest is the consequence that $\ham_0$ is, remarkably, Lorentz invariant~\cite{ko_passage_1978,hebenstreit_calculation_1979,Braud2025}. For our purpose an important consequence is that the dispersion relation of these degenerate modes is independent of the plasma velocity $\vec\beta$.


In fact, since the ray trajectory of a mode depends only on its dispersion relation~\cite{tracy_ray_2014}, that $\ham_0$ is independent of $\vec\beta$ then implies that the ray trajectory of these modes in a moving plasma is not affected by motion. Specifically, the ray equations of geometrical optics write $\dot{\vec x}=\uvec k$ and $\dot{\vec k}=0$ (see Appendix~\ref{app:pola_plane_redir}), implying $\ddot{\vec x}=0$, where an overdot indicates the total derivative with respect to the ray length ${s}$, and hats denote unitary vectors. The latter formally demonstrates that, as depicted in Fig.~\ref{fig:1}(a), the ray trajectory is straight, just like in a homogeneous plasma at rest. The ray equations further indicate that the group velocity $\vec v_g=d\vec x/dt\propto\dot{\vec x}$ is parallel to the wavevector $\vec k$, analogously to the behavior in an isotropic medium, and that it is constant along the ray. These results are consistent with the finding that a moving unmagnetized plasma does not bend light~\cite{ko_passage_1978}, as transverse drag remarkably vanishes when $n_g' n'=1$ with $n_g'$ the group index~\cite{Arnaud1976}.


The polarization plane and thus the spin, on the other hand, do vary along the ray due to motion. To see this, recall that the polarization of the degenerate modes lies in the plane normal to $\uvec w$. The electric spin density $\vec {\mathsf s}_{e}\propto \Im(\vec E^*\times\vec E)$~\cite{Berry2001,Bliokh2014,Neugebauer2018} is thus parallel to $\uvec w$. Eq.~\eqref{Eq:w_simplified} then shows unambiguously that the spin deviates from $\uvec k$, and thus from the ray tangent (see inset in Fig.~\ref{fig:1}(a)). This spin deviation is akin to the finite transverse spin~\cite{Bliokh2014,Aiello2015} found in bi-anisotropic media~\cite{Peng2019}, as a magneto-electric coupling is here introduced by motion~\cite{Minkowski1908,Kong2000}. Furthermore, the dependence of $\uvec w$ on $\vec \beta$ in Eq.~\eqref{Eq:w_simplified} reveals that the orientation of the spin with respect to the ray tangent will vary when the velocity varies along the ray, as shown in Fig.~\ref{fig:1}(a).



In short, the ray trajectory in a moving unmagnetized plasma is straight, yet motion introduces a tilt of the spin with respect to the ray tangent so that the polarization plane rotates as the velocity varies along the ray. 

\subsection{Spin-redirection Berry phase}
Let us now expose how this redirection of the polarization plane along the ray leads to the accumulation of a spin-redirection Berry phase~\cite{berry_interpreting_1987,bliokh_coriolis_2008,cohen_geometric_2019}. To do so, we consider as usual next-order corrections to geometrical optics~\cite{sluijter_general_2008,tracy_ray_2014,ruiz_first-principles_2015,ruiz_extending_2017,dodin_quasioptical_2019,onuki_quasi-local_2020,venaille_ray_2023}, working now with quasi-plane waves $\env_\pm e^{i\Phi}$. Here $\Phi=\int(\omega dt-\vec k\cdot d\vec x)$ is the usual dynamical phase, and the additional geometric phase $\psi$ manifests itself through the slow evolution of the envelope $\env_\pm=A_\pm e^{i\psi_{\pm}}\uvec\eta_{\pm}$ ~\cite{berry_quantal_1984,littlejohn_geometric_1991} with $\uvec \eta_{\pm}$ the mode unit polarization eigenvectors spanning the polarization plane, known up to a gauge choice.

By choosing these modes circularly polarized, we show in Appendix~\ref{app:Berry_phase} that the two modes are decoupled and thus do not exchange energy. As a result the transport equations for the Berry phases $\psi_{B_\pm}$ are independent, and the variation along the ray is simply given by~\cite{berry_quantal_1984,vinitskii_topological_1990,littlejohn_geometric_1991}
\begin{equation}
    \dot{\psi}_{B_\pm}=-\Im(\uvec\eta^\dag_\pm\dot{\uvec \eta}_\pm),
    \label{eq:evol_berry}
\end{equation}
clearly underlining that a redirection of the polarization plane, i.e. of the spin, will lead to slow phase accumulation. Classically, the eigenvectors $\uvec\eta$ are considered as functions of $(\vec k,\vec x)$, which leads to write Eq.~\eqref{eq:evol_berry} in terms of the Berry connections $\vec\cala^{(\vec k)}_\pm$ and $\vec\cala^{(\vec x)}_\pm$ associated respectively to the wavevector $\vec k$ and to the spatial coordinate $\vec x$~\cite{dodin_quasioptical_2019,perez_manifestation_2021}. These Berry connections describe the phase acquired by the eigenvectors with respect to variations of $\vec k$ or $\vec x$. However, we can instead consider $\uvec\eta$ as a function of $\vec w$, and accordingly define the Berry connection $\vec\cala^{(\vec w)}_\pm=\Im(\uvec\eta^\dag_\pm\vec(\nabla_{\vec w})\uvec\eta_\pm)$~\cite{braud_spin-orbit_2025}. Taking a constant gauge, we then find (see Appendix~\ref{app:Berry_phase}) that the Berry phase variation writes
\begin{equation}
    \dot{\psi}_{B_\pm}=-\dot{\vec w}\cdot\vec{\mathcal A}^{(\vec w)}_\pm.
    \label{eq:Berry_w}
\end{equation}
Importantly, we recognize that Eq.~\eqref{eq:Berry_w} is the exact analog of the Berry phase evolution known to arise from variations of the wavevector direction in static inhomogeneous isotropic medium~\cite{bliokh_geometrodynamics_2009,bliokh_spinorbit_2015}, if one replaces $\vec k$ by $\vec w$. We therefore demonstrate here that the redirection of $\vec w$ and thus of the electric spin density $\vec {\mathsf{s}}_e$ along a straight ray, here through variations of the velocity, lead to a Berry phase, just as a redirection of $\vec k$  leads to a Berry phase in inhomogeneous media~\cite{bliokh_geometrodynamics_2009,bliokh_spinorbit_2015}. Because in both cases the spin is redirected, both are spin-redirection effects. The redirection of $\vec w$ identified here can be thought of as a \emph{spin-deviation redirection}, in that it results from variations of the orientation of the spin with respect to the ray direction, whereas the redirection of $\vec k$ classically corresponds to a \emph{ray redirection}. 

The formal analogy identified here between a redirection of $\vec k$ and a redirection of $\vec w$ can be used further to offer an intuitive picture of the effect of spin deviation. To see this, let us define a virtual ray associated with our real ray, whose trajectory is tangent at all points to the vector $\uvec w$. This virtual trajectory is shown in Fig.~\ref{fig:1}(b) for the particular case of the dynamics of $\uvec w$ along the straight ray shown in Fig.~\ref{fig:1}(a). Through this construction, we get that the redirection of the polarization plane (i.e. of the spin $\vec {\mathsf{s}}_e$) by motion along the real ray is then exactly that due to the bending of the virtual trajectory. The spin-deviation-redirection phase induced by motion along a straight ray can thus be seen as the ray-redirection phase resulting from the associated nonplanar virtual trajectory. This equivalence can be explicitly demonstrated by considering the alternative gauge choice for which the eigenvectors $\uvec \eta_{\pm}$ are in the Frenet-Serret frame associated to $\vec w$. In fact as shown in Appendix~\ref{app:Berry_phase} Eq.~\eqref{eq:evol_berry} then gives $\dot{\psi}_{B\pm}=\pm\tau$ with $\tau=\ddot{\uvec w}\cdot(\uvec w\times\dot{\uvec w})/|\dot{\uvec w}|^2$, which we recognize as the torsion of the virtual trajectory, just like the Berry phase evolution is given by the torsion of the real ray in static isotropic media~\cite{berry_interpreting_1987,bliokh_geometrodynamics_2009}.



\subsection{Polarization rotation from spin redirection}
Let us finally show how this spin-redirection Berry phase affects the wave polarization. To do so we define the wave polarization unit vector $\uvec e=\vec Ee^{-i\Phi}/|\vec E|$ which captures the wave polarization state and direction, with $\vec E=\vec E_++\vec E_-$ the total wave field obtained by summing the two modes. The transport of $\uvec e$ is then given by the evolution of the slow phases $\dot{\psi}_{\pm}$ and of the eigenvectors $\dot{\uvec \eta}_{\pm}$~\cite{braud_spin-orbit_2025}.


\begin{figure}
    \centering
    \includegraphics[width=\linewidth]{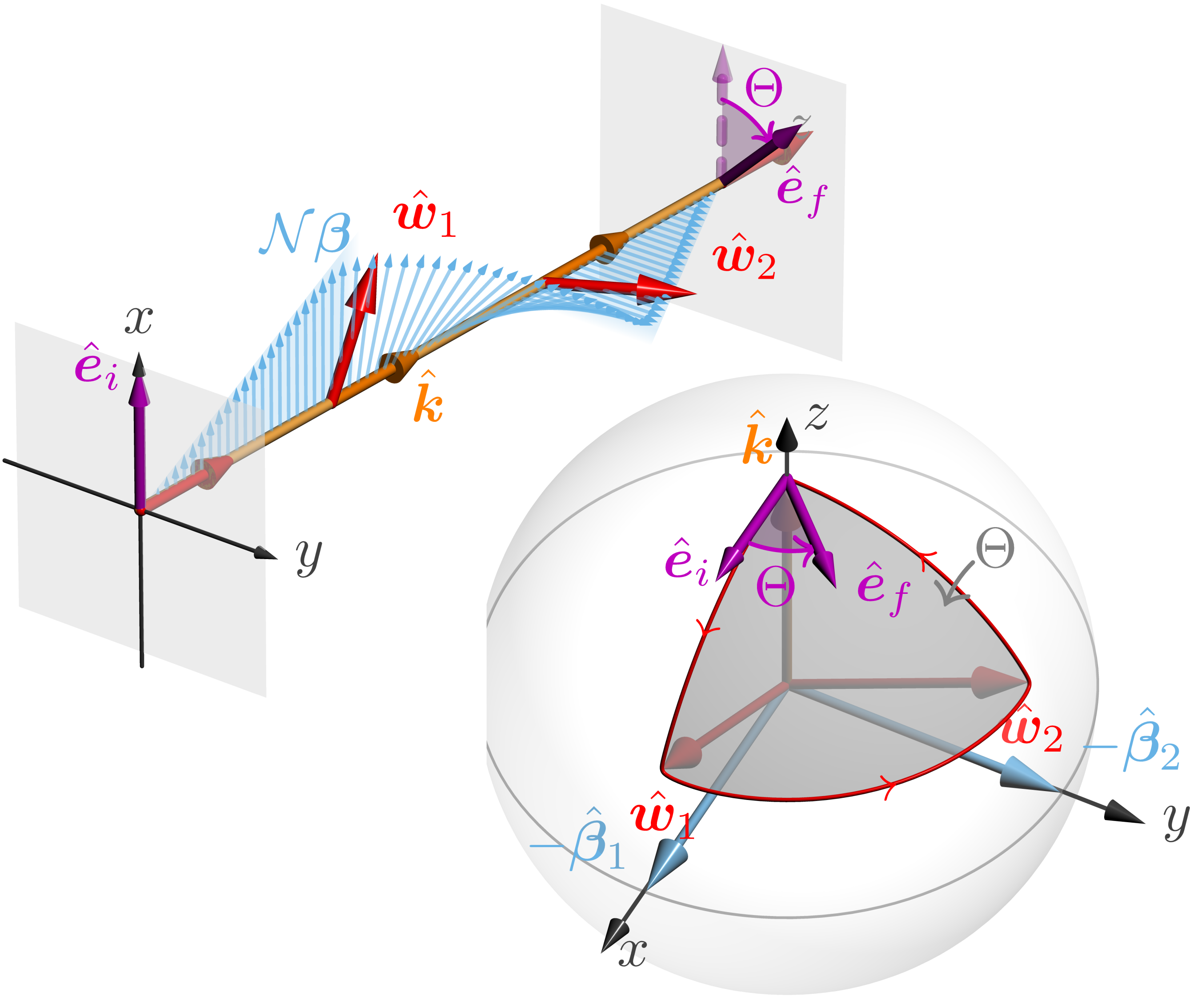}
    \begin{picture}(0,0)
        \put(-0.51\linewidth,7.34cm){\textbf{(a)}} 
    \end{picture}
    \begin{picture}(0,0)
        \put(3.7cm,4.8cm){\textbf{(b)}} 
    \end{picture}
    \caption{(a) Cyclic evolution of the velocity ($\vec 0\rightarrow\uvec x\rightarrow\uvec y\rightarrow \vec 0$) along a straight ray and associated rotation of the wave polarization $\uvec e$ from $\uvec e_i$ to $\uvec e_f$. (b) The polarization rotation angle $\Theta$, given by the accumulated spin-deviation-redirection Berry phase $\psi_B$, is equal to the solid angle enclosed by the closed contour described by $\uvec w$ around $\vec v_g$, analogously to the polarization rotation typical of a redirection of $\vec k$.}
    \label{fig:2}
\end{figure}
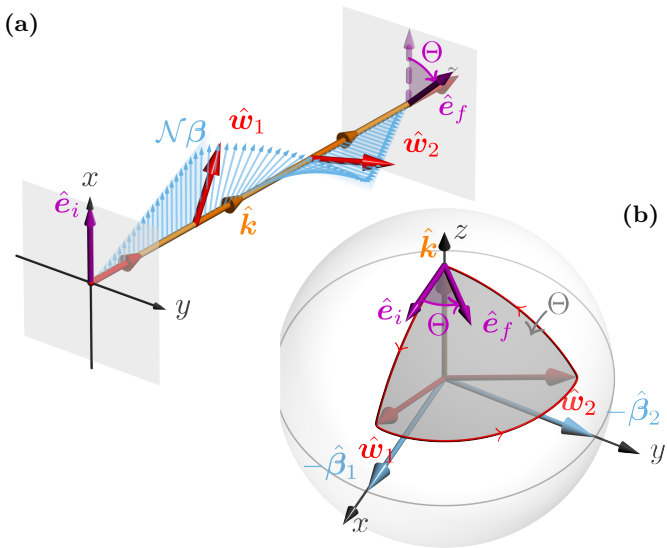

Looking specifically at the effect of the spin-deviation-redirection phase due here to motion, we show in Appendix~\ref{app:pola_rotation} that it acts on the polarization in a way much similar to the parallel transport along nonplanar rays in static media. Precisely, we find that its contribution to $d\uvec e/d{s}$ can be written as the parallel transport $-(\dot{\uvec w}\cdot\uvec e)\uvec w$, which then manifests as a global rotation of the wave polarization $\uvec e$ in the polarization plane normal to $\uvec w$. This is analogous to Rytov's law and Rytov's rotation~\cite{rytov_sur_1938, vinitskii_topological_1990, bliokh_geometrodynamics_2009}, except that the polarization is here parallel transported along a path with tangent vector $\uvec w$, rather than along the real (here straight) ray~\cite{rytov_sur_1938,berry_interpreting_1987, bliokh_geometrodynamics_2009}. This path is, however, precisely the virtual trajectory defined above. The effect of the spin-deviation-redirection phase on polarization is thus exactly that of a parallel transport along the virtual trajectory.

Recognizing that parallel transport along the virtual trajectory corresponds to parallel transport on the $\uvec w$-sphere
makes it possible to draw a familiar picture of the effect on polarization of a cyclic evolution of $\uvec w$ (i.e. of $\vec \beta$). Specifically, by considering as illustrated in Fig.~\ref{fig:2} a closed contour for $\uvec w$, the Berry phase $\psi_B$ is independent of the gauge~\cite{samuel_general_1988,cohen_geometric_2019}, and the polarization rotation angle $\Theta$ is given by the solid angle enclosed by the contour (see Appendix~\ref{app:pola_rotation}). This polarization rotation along a straight ray manifests the holonomy of the Berry connection on the $\uvec w$-sphere. This representation unveils how the effect of the spin-deviation-redirection phase on polarization is geometrically equivalent to the well-known parallel transport of polarization on the $\uvec k$-sphere due to the ray-redirection phase accumulated along nonplanar rays~\cite{bliokh_geometrodynamics_2009,bliokh_spinorbit_2015,Tomita1986,berry_interpreting_1987}. 

Note finally for completeness that, as noted for non-dispersive media~\cite{braud_spin-orbit_2025}, the dynamics of the polarization $\uvec e$ in a moving medium is not purely the effect of the parallel transport due to the spin-redirection Berry phase. Specifically, a nonuniform velocity field leads to an additional drag contribution, long predicted~\cite{player_dragging_1976} and observed~\cite{jones_rotary_1976} in rotating dielectrics, and more recently studied in plasmas~\cite{gueroult_determining_2019,Langlois2024,Gueroult2025}.

\subsection{Example}
\begin{figure}
    \centering
    \includegraphics[width=\linewidth]{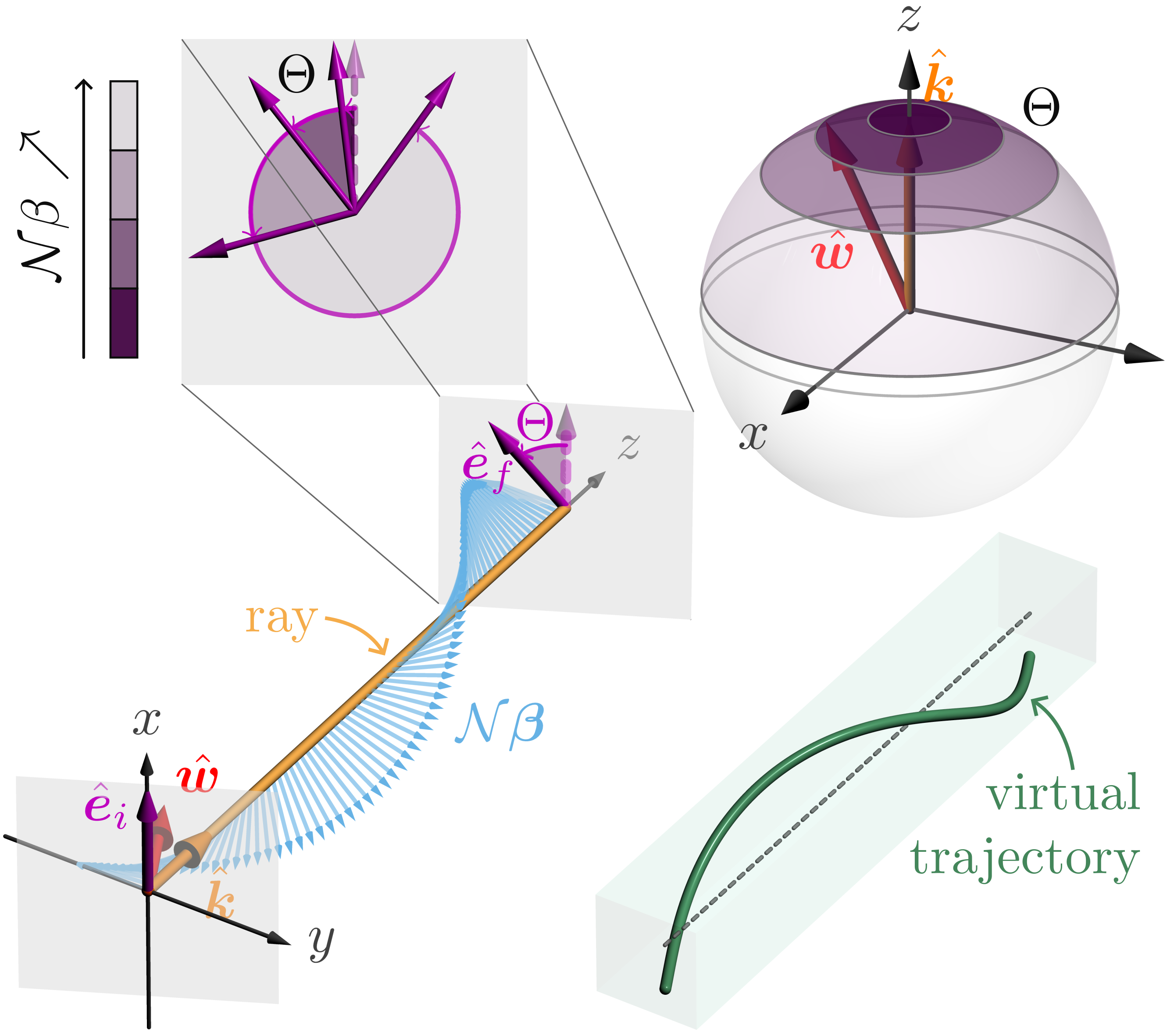}
    \begin{picture}(0,0)
        \put(-0.495\linewidth,7.71cm){\textbf{(a)}} 
    \end{picture}
    \begin{picture}(0,0)
        \put(0.42\linewidth,7.71cm){\textbf{(b)}} 
    \end{picture}
    \begin{picture}(0,0)
        \put(0.42\linewidth,3.875cm){\textbf{(c)}} 
    \end{picture}
    \caption{(a) Rotation of polarization due to a rotation of the velocity along the ray. (b) Cone described by $\uvec w$ around $\vec v_g\parallel \vec k$ during the cycle. (c) Associated helix virtual trajectory. The solid angle described by $\uvec w$ grows with $\mathcal N \beta$ (darker to lighter colors), leading to significant polarization rotation for $\mathcal{N}\beta=\mathcal{O}(1)$. }
    \label{fig:3}
\end{figure}
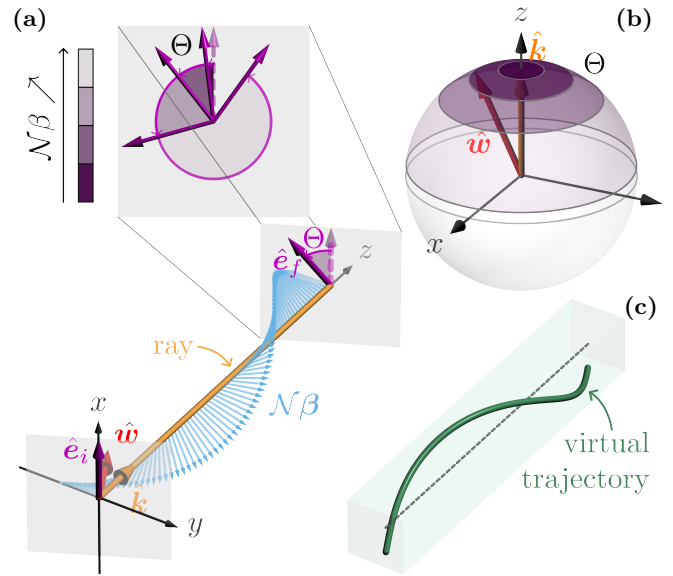

To further expose the formal analogy with nonplanar rays and quantify this spin-deviation-redirection effect, let us finally consider as an example the case where the virtual trajectory is a helix, as studied by Bliokh for light propagating in a gradient index medium~\cite{bliokh_geometrodynamics_2009}. For that consider as shown in Fig.~\ref{fig:3}(a) a velocity field $\vec \beta$ that is purely transverse to the constant wavevector $\vec k$, and that uniformly rotates along the ray, which we explicitly write $\vec\beta({s})=e^{{s}\Omega\utensor k}\vec\beta_0$. In this case $\uvec w$, which is dragged by motion, describes a cone as shown in Fig.~\ref{fig:3}(b), and we verify that the virtual trajectory is then a helix with constant curvature and torsion $\tau=\Omega/\sqrt{1+\mathcal{N}^2\beta^2}$ as depicted in Fig.~\ref{fig:3}(c).

Considering first the effect of the nonplanar virtual trajectory, it has been shown that for a helix polarization rotation is simply a function of the ray torsion~\cite{bliokh_geometrodynamics_2009}, with $\Theta=2\pi\left(1-\tau/\Omega\right)$. Considering alternatively the effect of the spin-deviation-redirection phase, since $\vec \beta$ is here normal to $\vec k$, expanding Eq.~\eqref{Eq:w_simplified} simply gives $\vec w\propto \uvec k+\mathcal{N}\vec\beta$, from which we easily verify that $\Theta$ indeed corresponds to the solid angle of the cone drawn by $\uvec w$ during a cycle. Thus, both pictures give, as anticipated, the same result.

Incidentally, this simple scaling can be used to infer the importance of this polarization rotation effect. For $\mathcal{N}\beta\ll1$, the deviation of the polarization plane in Eq.~\eqref{Eq:w_simplified} is small, or equivalently the virtual trajectory is nearly straight, and the polarization is therefore barely affected. On the other hand, as shown in Fig.~\ref{fig:3}(b), the solid angle described by $\uvec w$ over a cycle grows with $\mathcal{N}\beta$, leading to large polarization rotation $\Theta$ for $\mathcal{N}\beta=\mathcal{O}(1)$. This shows that significant effects can arise not only from large velocities, but also at lower $\beta$ for sufficiently large Fresnel drag coefficients $\mathcal{N}\gg1$. In a plasma the latter notably occurs for a wave frequency close to the plasma frequency $\omega_p$ as $n'(\omega)\to 0$. Note finally that while the apparent rotation after one cycle goes to zero when $\mathcal{N}\beta\to \infty$, the polarization is in this case fully dragged, leading to significant polarization effects even on a fraction of cycle.

\section{Discussion}
The choice to work here with a moving unmagnetized plasma has been motivated by the fact that rays are then straight, showing decisively how spin-orbit redirection effects do not require, contrary to common wisdom, nonplanar rays. However, it is important to stress that, since it simply requires a redirection of the spin with respect to the ray tangent, this effect is expected to be at play in a range of systems much broader than moving media. Importantly, we note that a tilt of the spin with respect to the ray tangent is in particular intrinsic to transverse spin (T-spin) waves~\cite{Bliokh2014,Aiello2015}, suggesting that a spin-deviation-redirection phase may arise in systems supporting T-spin.


A particular example is plane waves in static magnetized plasmas, for which the spin $\vec {\mathsf{s}}_e$ is generally neither purely transverse nor purely longitudinal~\cite{Booker1984}, with an inclination of the spin with respect to the wavevector that depends on the local plasma parameters and the wave frequency. Finite T-spin is in fact more generally characteristic of gyrotropic media~\cite{Gong2021}, as well as of bianisotropic media~\cite{Peng2019}. Beyond bulk media, finite T-spin has also been demonstrated in structured optical fields such as evanescent waves~\cite{Bliokh2014}, strongly focused beams~\cite{Neugebauer2015}, two-waves interference~\cite{Bekshaev2015}, and plasmons~\cite{CanaguierDurand2014,Bliokh2012}. In all of these systems, the finite transverse spin component will in general lead to a deviation of $\vec {\mathsf{s}}_e$ with respect to the ray tangent, creating the conditions for the new spin-redirection mechanism exposed here. This broader applicability in turn suggests that spin-deviation-redirection SOIs could be more widely at play in nature, possibly adding to the recent discovery that SOIs play an important role in space~\cite{Leclerc2025} and astrophysics~\cite{Zhang2024,Oancea2024}. In addition, because the inclination of the spin with respect to the wavevector can, as demonstrated in gyrotropic media~\cite{Gong2021} and focused light~\cite{Bliokh2015a,Chen2017}, be continuously varied from purely longitudinal to purely transverse, the new spin-redirection contribution identified here offers a new degree of freedom to control SOIs, besides the known ray-redirection effects~\cite{bliokh_geometrodynamics_2009,Fu2024}. This effect could in turn be used to manipulate waves in novel ways, complementing notably the strategies utilizing wavelength-scale light-matter interactions in metamaterials~\cite{Cardano2015,Ling2017,jisha_geometric_2021}.

\section{Summary}
Our work shows that a spin-orbit interaction of spin-redirection type can occur even if the ray trajectory is planar. This is made possible by the fact that the electric spin density, which prescribes the orientation of the polarization plane, is not necessarily parallel to the group velocity. As a result the polarization plane can be redirected independently of the ray direction. This polarization plane redirection then classically leads to a spin-redirection Berry phase, which in turn manifests itself as a rotation of polarization. We expose here this phenomenon through the example of a moving unmagnetized plasma where rays are singularly straight, thus unambiguously revealing the existence of an additional mechanism for the spin-redirection phase. Because it stems from a deviation of the spin with respect to the ray tangent, we refer to it as spin-deviation redirection. By identifying a connection with transverse spin waves, we finally show that this effect should be at play in a much larger class of systems, and that it will in general add to the usual ray-redirection effects induced by nonplanar rays. Given the broad and active interest in SOIs, the identification here of a new mechanism for SOIs could enable to the discovery of a new class of light manipulation schemes, but also of the signature of SOIs in a wider range of natural phenomena. 

\section*{Acknowledgments}

This work is supported by the Agence Nationale de la Recherche (ANR) through the WaRP project (ANR-21-CE30-0002). This work has been carried out within the framework of the EUROfusion Consortium, via the Euratom Research and Training Programme (Grant Agreement No 101052200 – EUROfusion). Views and opinions expressed are however those of the authors only and do not necessarily reflect those of the European Union or the European Commission. 

The authors thank J. Langlois for constructive discussions.

\section*{Data availability}

No data were created or analyzed in this study.

\appendix

\section{Polarization plane redirection}\label{app:pola_plane_redir}
\noindent
\textit{Dispersion matrix.}-- Following Refs.~\cite{anile_relativistic_2005,gedalin_waves_2001}, electromagnetism in a moving plasma can be described by the covariant Maxwell equations $\partial_\nu F^\munu=\mu_0 j^\mu$ with $F^\munu$ the Faraday tensor and $j^\mu$ the four-current, together with the covariant fluid equation $\partial_\nu T^{\nu\mu}_s=F^\mu_\nu j_s^\nu$ for each species $s$ with $T^{\nu\mu}_s=n_sm_su_s^\nu u_s^\mu$ the stress-energy tensor and $j^\nu_s=n_sq_su^\nu_s$. Here $n_s$, $m_s$, $q_s$ and $u_s^\mu=(\gamma_s c,\gamma_s \vec v_s)$ are respectively the density, the mass, the charge and the four-velocity for species $s$, and $j^\mu=\sum_s j^\mu_s$.

Looking for  plane  wave  solutions with a phase $e^{i k_\mu x^\mu}$, one can then obtain as done in Refs.~\cite{tam_covariant_1968,melrose_covariant_1973} a $0^\textrm{th}$ order wave equation for the  perturbation of the Faraday tensor $F_1^\munu$, and from there for the four-potential perturbation $A_1^\mu$ of the form $D^{\mu}_\nu A_1^\nu=0$ with
\begin{multline}
\label{Eq:disp_A}
    D^{\mu}{}_\nu=k_\lambda k^\lambda\eta^\mu_\nu-k^\mu k_\nu-\frac{\omega_p^2}{(k_\lambda\beta^\lambda)^2}\big[(k_\lambda\beta^\lambda)^2\eta^\mu_\nu\\-(k_\lambda\beta^\lambda)\left(k^\mu \beta_\nu+\beta^\mu k_\nu\right)+k_\lambda k^\lambda \beta^\mu\beta_\nu\big].
\end{multline}
Using finally the definition of the four-vector electric field $E^\mu=F_1^\munu \beta_\nu$, so that $E^\mu=\left(k^\mu \beta_\nu-(k_\lambda k^\lambda) \eta^\mu_\nu\right) A_1^\nu$, Eq.~\eqref{Eq:disp_A} gives the dispersion matrix for the spatial part of the electric field $\vec E$
\begin{equation}\label{eq:exact_dispersion_matrix_E}
    \tensor D=\ham_0\left(\eye-\frac{\gamma^2\omega^2}{\omega'^2}(n'^2(\omega)-1)\vec\beta\vec\beta^T\right)+\vec w\vec w^T.
\end{equation}
Here $\ham_0=(\omega^2-\omega_p^2)/c^2-k^2$ and $n'(\omega)^2=1-\omega_p^2/\omega^2$ are respectively the dispersion function and the refractive index of the plasma at rest with $\omega_p^2=\omega_{pe}^2+\omega_{pi}^2$ the plasma angular frequency and $\omega'=\gamma(\omega-\vec k\cdot\vec\beta c)$ the Doppler shifted frequency, and 
\begin{equation}
    \vec w=\vec k+\frac{\gamma\omega^2}{\omega'c} (n'^2(\omega)-1)\vec\beta(\vec x).
\end{equation}
Eq.~\eqref{eq:exact_dispersion_matrix_E}, which we verify reduces to Ref.~\cite{lee_radiation_1966} in the unmagnetized limit, gives Eqs.~\eqref{Eq:D0_simplified} and \eqref{Eq:w_simplified}, and by identification the definitions of $\tensor M$, $\alpha$ and $\mathcal{N}$.

\noindent\textit{Ray equations.}-- The ray equations are classically the equations of motion for the trajectory of waves described by rays in the framework of geometrical optics~\cite{tracy_ray_2014}. They read $\dot {\vec x}=-\vec\nabla_{\vec k}\ham_0/|\vec\nabla_{\vec k}\ham_0|$, $\dot t=\partial_\omega\ham_0/|\vec\nabla_{\vec k}\ham_0|$, $\dot{\vec k}=\vec\nabla_{\vec x}\ham_0/|\vec\nabla_{\vec k}\ham_0|$ and $\dot\omega=-\partial_t\ham_0/|\vec\nabla_{\vec k}\ham_0|$ where a dot indicates a derivative with respect to the ray parameter ${s}$ taken here to be the physical length of the ray. Using $\ham_0$ as given above we find $\dot{\vec x}=\uvec k$, $\dot t=1/(n'c)$, $\dot{\vec k}=0$ and $\dot\omega=0$, as indicated in the main text.


\section{Spin-redirection Berry phase}\label{app:Berry_phase}
\noindent

\noindent\textit{Degenerate propagative modes.}-- From Eq.~\eqref{eq:exact_dispersion_matrix_E} we get
\begin{equation}\label{eq:exact_det}
    \det\tensor D=\frac{\omega^2}{c^2}\left(1+(n'^2-1)\frac{\omega^2}{\omega'^2}\right)\ham_0^2,
\end{equation}
so that $\ham_0$ is an eigenvalue of multiplicity $2$. Because the dispersion matrix reduces to the projection matrix along $\vec w$ when $\ham_0$ vanishes, the two degenerate modes associated with this eigenvalue must lay in the plane normal to $\uvec w$. We can then choose as a basis of this nullspace of dimension 2 the right and left circular polarizations defined by $\uvec \eta_\pm=\vec\eta_\pm/|\vec \eta_\pm|$ with
\begin{equation}
    \vec \eta_\pm=\left(\eye-\uvec w\uvec w^T\pm i\utensor w\right)\uvec u
\end{equation}
where $\uvec u$ is a gauge vector field that can be taken arbitrarily, as long as it is not parallel to $\uvec w$. 

Note that the fact that $\ham_0$ is an eigenvalue of multiplicity $2$ is consistent with the fact that the dispersion functions of the modes propagating in the moving medium can be obtained by performing Lorentz transformations in the corresponding dispersion functions of the medium at rest~\cite{censor_dispersion_1980,mccall_relativity_2007,censor_relativistic_2010}, and that the dispersion relation of an unmagnetized plasma is singularly Lorentz invariant~\cite{ko_passage_1978}. For completeness we verify that the bracketed term in Eq.~\eqref{eq:exact_det} does match the dispersion relation obtained through Lorentz transformations of the dispersion relation of the electrostatic mode in rest frame $\omega'^2-\omega_p^2=0$.
\\

\noindent\textit{Berry phase accumulation.}-- We adopt here the extended geometrical optics framework derived in Refs.~\cite{ruiz_extending_2017,dodin_quasioptical_2019}. Consistent with this model and assuming that the amplitude of the third mode is negligible compared to the amplitudes of the two degenerate modes we are interested in, we write $\vec E(\vec x,t)=A(\vec x,t)\tensor\Xi(\vec x,t)\uvec \xi(\vec x,t) e^{i\Phi(\vec x,t)}$ the total wave field. Here $A$ is the amplitude of the total wave field, $\Phi=\int (\omega dt-\vec k\cdot d\vec x)$ is the dynamical phase~\cite{berry_quantal_1984,venaille_ray_2023}, $\tensor\Xi=(\uvec \eta_+,\uvec \eta_-)$ is the polarization matrix~\cite{dodin_quasioptical_2019} and $\uvec \xi=(\varkappa_+e^{i\psi_+}, \varkappa_-e^{i\psi_-})^T$ is the Jones vector containing the normalized complex amplitudes of the modes with $\psi_\pm$ the phases of the envelope. The latter, which include the Berry phases $\psi_{B\pm}$, are referred to as \emph{slow phases} as they are assumed to vary slowly compared to the dynamical phase ~\cite{ruiz_first-principles_2015,ruiz_lagrangian_2015}.

Because the circularly polarized modes conveniently verify the property $\uvec\eta_+^*=\uvec\eta_-$, the spin-orbit Hamiltonian turns out to be diagonal. The rate of change of the Berry phase is then simply $\dot\psi_{B_\pm}=-\Im(\uvec \eta_\pm^\dag\dot{\uvec\eta}_\pm)$~\cite{vinitskii_topological_1990,littlejohn_geometric_1991}. Since the eigenvectors are functions of $(\vec w,\vec u)$, we can write $\dot\psi_{B_\pm}=-\dot{\vec w}\cdot\vec{\mathcal A}^{(\vec w)}_\pm-\dot{\vec u}\cdot\vec{\mathcal A}^{(\vec u)}_\pm$ where we have defined $\vec{\mathcal A}^{(\vec w)}_\pm=\Im(\uvec \eta^\dag_\pm(\vec\nabla_{\vec w})\uvec\eta_\pm)$ and $\vec{\mathcal A}^{(\vec u)}_\pm=\Im(\uvec \eta^\dag_\pm(\vec\nabla_{\vec u})\uvec\eta_\pm)$ the Berry connections with respect to respectively $\vec w$~\cite{braud_spin-orbit_2025} and $\vec u$. This result then immediately yields Eq.~\eqref{eq:Berry_w} for a constant gauge vector $\uvec u$.

Looking for an explicit form, we find after some algebra that $\vec{\mathcal A}^{(\vec w)}_\pm=\mp(\uvec w\cdot\uvec u)\uvec u\times\uvec w/[1-(\uvec w\cdot\uvec u)^2]/|\vec w|$ and $\vec{\mathcal A}^{(\vec u)}_\pm=\pm\uvec u\times\uvec w/[1-(\uvec w\cdot\uvec u)^2]/|\vec u|$, giving finally the rate of change of the Berry phase
\begin{equation}\label{eq:dot_psi_B}
    \dot{\psi}_{B_\pm}=\pm\frac{(\uvec w\cdot\uvec u)\dot{\vec w}/|\vec w|-\dot{\vec u}/|\vec u|}{1-(\uvec w\cdot\uvec u)^2}\cdot\left(\uvec u\times\uvec w\right).
\end{equation}
Noting that $\dot\psi_{B_+}=-\dot\psi_{B_-}$, we write $\dot\psi_{B}=|\dot\psi_{B_+}|$.
\\

\noindent\textit{Relation to the torsion.}-- Taking finally a gauge $\uvec u=\dot{\uvec w}/|\dot{\uvec w}|$, which corresponds to defining the circularly polarized modes in the Frenet-Serret frame as done for instance in Ref.~\cite{bliokh_geometrodynamics_2009}, Eq.~\eqref{eq:dot_psi_B} confirms that $\dot{\psi}_{B_\pm}=\pm\tau$ with $\tau=\ddot{\uvec w}\cdot(\uvec w\times\dot{\uvec w})/|\dot{\uvec w}|^2$ the local torsion of the virtual trajectory. This result is analogous to that found in Refs.~\cite{berry_interpreting_1987,vinitskii_topological_1990,bliokh_geometrodynamics_2009} except that in these references the torsion is that of the real ray.

\section{Polarization rotation}\label{app:pola_rotation}
\noindent
\noindent\textit{Parallel transport law for $\uvec e$.}-- Given our definition for the total wave field, the polarization unit vector is
\begin{equation}
    \uvec e=\tensor\Xi\uvec \xi=\varkappa_+\uvec\eta_+e^{i\psi_+}+\varkappa_-\uvec\eta_-e^{i\psi_-}.
\end{equation}
According to Ref.~\cite{littlejohn_geometric_1991}, it is gauge independent and is therefore a physical observable that characterizes the polarization state and direction.

Taking the derivative of $\uvec e$ with respect to the ray length ${s}$, we get a polarization transport equation of the form $\dot{\uvec e}=\vec t+\vec d$ where 
\begin{equation}
    \vec t=(\dot{\tensor \Xi}\tensor\Xi^\dag+i\sum_{m=\pm}\dot\psi_{Bm}\uvec\eta_m\uvec\eta_m^\dag)\uvec e
\end{equation}
is the SOI contribution and $\vec d$ is a supplemental mechanical drag contribution that stems from motion, as shown in Ref.~\cite{braud_spin-orbit_2025}. Remarking that $\vec t=(\dot{\tensor \Xi}\tensor\Xi^\dag-\tensor\Xi\tensor\Xi^\dag\dot{\tensor\Xi}\tensor\Xi^\dag)\uvec e=(\eye-\tensor\Xi\tensor\Xi^\dag)d(\tensor\Xi\tensor\Xi^\dag)/d{s}\uvec e$ and that $\tensor\Xi\tensor\Xi^\dag=\eye-\uvec w\uvec w^T$ is the projector on the plane normal to $\uvec w$, we find that $\vec t$ can be written in the form of the parallel transport law $\vec t=-\uvec w\dot{\uvec w}^T\uvec e$~\cite{vinitskii_topological_1990,bliokh_geometrodynamics_2009}.
\\

\noindent\textit{Relation to the solid angle.}-- For circularly polarized modes as considered here the slow phase leads to a phase-shift which then manifests as a rotation of the polarization. Because these modes rotate in opposite directions, and because $\dot\psi_{B_+}=-\dot\psi_{B_-}$, the Berry phase difference accumulated between the modes is $2\int_s\dot\psi_B d{s}$. For a cyclic evolution this Berry phase difference is independent of the gauge~\cite{samuel_general_1988,cohen_geometric_2019}, and we write it here $2\psi_B$. Because our total wave field is the sum of these modes, this phase difference between circularly polarized modes then leads to a rotation of the wave polarization $\uvec e$ by an angle $\psi_B$ modulo $2\pi$. In other words $\Theta=\psi_B[2\pi]$.

Meanwhile, the solid angle enclosed by the contour described by $\uvec w$ on the unit $\uvec w$-sphere is $\varOmega=\iint_{\mathcal S}\uvec w/w^2d^2\vec w$. Interestingly, we find from the expression of $\vec{\mathcal A}^{(\vec w)}_\pm$ given above that the integrand $\uvec w/w^2$ is precisely the Berry curvature with respect to $\vec w$, i.e. $\vec F^{(\vec w)}_\pm=\vec\nabla_{\vec w}\times\vec\cala^{(\vec w)}_\pm=\pm\uvec w/w^2$. This result is again analogous to the one known in static non-dispersive media~\cite{bliokh_geometrodynamics_2009}, except that we consider here the solid angle described by $\uvec w$ instead of $\uvec k$. From Stokes' theorem, we then have $\varOmega=\pm\oint_{\mathcal C}\dot{\vec w}\cdot\vec\cala^{(\vec w)}_\pm d{s}$, where we recognize from Eq.~\eqref{eq:Berry_w} that the integrand is simply $\dot\psi_B$. This shows that the solid angle $\varOmega$ is also equal to the Berry phase accumulated during the cycle $\psi_B$. 

Putting these pieces together confirms that the rotation of polarization over one cycle is indeed equal to the solid angle enclosed by the contour described by $\uvec w$ on the unit sphere, i.e. $\varOmega=\Theta$.

\bibliography{references,refs}

@article{littlejohn_geometric_1991,
	title = {Geometric phases in the asymptotic theory of coupled wave equations},
	volume = {44},
	issn = {1050-2947, 1094-1622},
	url = {https://link.aps.org/doi/10.1103/PhysRevA.44.5239},
	doi = {10.1103/PhysRevA.44.5239},
	number = {8},
	urldate = {2024-02-14},
	journal = {Phys. Rev. A},
	author = {Littlejohn, Robert G. and Flynn, William G.},
	month = oct,
	year = {1991},
	pages = {5239--5256},
}

@article{bliokh_coriolis_2008,
	title = {Coriolis {Effect} in {Optics}: {Unified} {Geometric} {Phase} and {Spin}-{Hall} {Effect}},
	volume = {101},
	issn = {0031-9007, 1079-7114},
	shorttitle = {Coriolis {Effect} in {Optics}},
	url = {https://link.aps.org/doi/10.1103/PhysRevLett.101.030404},
	doi = {10.1103/PhysRevLett.101.030404},
	number = {3},
	urldate = {2024-02-14},
	journal = {Phys. Rev. Lett.},
	author = {Bliokh, Konstantin Y. and Gorodetski, Yuri and Kleiner, Vladimir and Hasman, Erez},
	month = jul,
	year = {2008},
	pages = {030404},
}

@article{bliokh_geometrodynamics_2009,
	title = {Geometrodynamics of polarized light: {Berry} phase and spin {Hall} effect in a gradient-index medium},
	volume = {11},
	issn = {1464-4258, 1741-3567},
	shorttitle = {Geometrodynamics of polarized light},
	url = {https://iopscience.iop.org/article/10.1088/1464-4258/11/9/094009},
	doi = {10.1088/1464-4258/11/9/094009},
	number = {9},
	urldate = {2024-02-14},
	journal = {J. Opt. A: Pure Appl. Opt.},
	author = {Bliokh, Konstantin Y},
	month = sep,
	year = {2009},
	pages = {094009},
}

@article{bliokh_spinorbit_2015,
	title = {Spin–orbit interactions of light},
	volume = {9},
	issn = {1749-4885, 1749-4893},
	url = {https://www.nature.com/articles/nphoton.2015.201},
	doi = {10.1038/nphoton.2015.201},
	number = {12},
	urldate = {2024-02-14},
	journal = {Nat. Photonics},
	author = {Bliokh, K. Y. and Rodríguez-Fortuño, F. J. and Nori, F. and Zayats, A. V.},
	month = dec,
	year = {2015},
	pages = {796--808},
}

@article{cohen_geometric_2019,
	title = {Geometric phase from {Aharonov}–{Bohm} to {Pancharatnam}–{Berry} and beyond},
	volume = {1},
	issn = {2522-5820},
	url = {https://www.nature.com/articles/s42254-019-0071-1},
	doi = {10.1038/s42254-019-0071-1},
	number = {7},
	urldate = {2024-02-14},
	journal = {Nat. Rev. Phys.},
	author = {Cohen, Eliahu and Larocque, Hugo and Bouchard, Frédéric and Nejadsattari, Farshad and Gefen, Yuval and Karimi, Ebrahim},
	month = jun,
	year = {2019},
	pages = {437--449},
}

@article{jisha_geometric_2021,
	title = {Geometric {Phase} in {Optics}: {From} {Wavefront} {Manipulation} to {Waveguiding}},
	volume = {15},
	issn = {1863-8880, 1863-8899},
	shorttitle = {Geometric {Phase} in {Optics}},
	url = {https://onlinelibrary.wiley.com/doi/10.1002/lpor.202100003},
	doi = {10.1002/lpor.202100003},
	number = {10},
	urldate = {2024-02-14},
	journal = {Laser Photonics Rev.},
	author = {Jisha, Chandroth Pannian and Nolte, Stefan and Alberucci, Alessandro},
	month = oct,
	year = {2021},
	pages = {2100003},
}

@article{perez_manifestation_2021,
	title = {Manifestation of the {Berry} curvature in geophysical ray tracing},
	volume = {477},
	issn = {1364-5021, 1471-2946},
	doi = {10.1098/rspa.2020.0844},
	number = {2248},
	urldate = {2024-02-14},
	journal = {Proc. R. Soc. A},
	author = {Perez, N. and Delplace, P. and Venaille, A.},
	year = {2021},
	pages = {20200844},
}

@article{sluijter_general_2008,
	title = {General polarized ray-tracing method for inhomogeneous uniaxially anisotropic media},
	volume = {25},
	issn = {1084-7529, 1520-8532},
	url = {https://opg.optica.org/abstract.cfm?URI=josaa-25-6-1260},
	doi = {10.1364/JOSAA.25.001260},
	number = {6},
	urldate = {2024-02-14},
	journal = {J. Opt. Soc. Am. A},
	author = {Sluijter, Maarten and De Boer, Dick K. G. and Braat, Joseph J. M.},
	month = jun,
	year = {2008},
	pages = {1260},
}

@article{ruiz_first-principles_2015,
	title = {First-principles variational formulation of polarization effects in geometrical optics},
	volume = {92},
	issn = {1050-2947, 1094-1622},
	url = {https://link.aps.org/doi/10.1103/PhysRevA.92.043805},
	doi = {10.1103/PhysRevA.92.043805},
	number = {4},
	urldate = {2024-02-14},
	journal = {Phys. Rev. A},
	author = {Ruiz, D. E. and Dodin, I. Y.},
	month = oct,
	year = {2015},
	pages = {043805},
}

@article{venaille_ray_2023,
	title = {From ray tracing to waves of topological origin in continuous media},
	volume = {14},
	issn = {2542-4653},
	url = {https://scipost.org/10.21468/SciPostPhys.14.4.062},
	doi = {10.21468/SciPostPhys.14.4.062},
	number = {4},
	urldate = {2024-02-14},
	journal = {Scipost Phys.},
	author = {Venaille, Antoine and Onuki, Yohei and Perez, Nicolas and Leclerc, Armand},
	month = apr,
	year = {2023},
	pages = {062},
}

@article{ruiz_extending_2017,
	title = {Extending geometrical optics: {A} {Lagrangian} theory for vector waves},
	volume = {24},
	doi = {10.1063/1.4977537},
	number = {5},
	journal = {Phys. Plasma},
	author = {Ruiz, D. E. and Dodin, I. Y.},
	year = {2017},
	pages = {055704}
}

@article{onuki_quasi-local_2020,
	title = {Quasi-local method of wave decomposition in a slowly varying medium},
	volume = {883},
	issn = {0022-1120, 1469-7645},
	url = {https://www.cambridge.org/core/product/identifier/S0022112019008255/type/journal_article},
	doi = {10.1017/jfm.2019.825},
	urldate = {2024-02-14},
	journal = {J. Fluid Mech.},
	author = {Onuki, Yohei},
	month = jan,
	year = {2020},
	pages = {A56},
}

@article{lee_radiation_1966,
	title = {Radiation in a {Moving} {Anisotropic} {Medium}},
	volume = {1},
	issn = {0048-6604, 1944-799X},
	url = {https://agupubs.onlinelibrary.wiley.com/doi/10.1002/rds196613313},
	doi = {10.1002/rds196613313},
	number = {3},
	urldate = {2024-02-14},
	journal = {Radio Sci.},
	author = {Lee, S. W. and Lo, Y. T.},
	month = mar,
	year = {1966},
	pages = {313--324},
}

@article{melrose_covariant_1973,
	title = {A covariant formulation of wave dispersion},
	volume = {15},
	issn = {0032-1028},
	url = {https://iopscience.iop.org/article/10.1088/0032-1028/15/2/002},
	doi = {10.1088/0032-1028/15/2/002},
	number = {2},
	urldate = {2024-02-14},
	journal = {Plasma Phys.},
	author = {Melrose, D B},
	month = feb,
	year = {1973},
	pages = {99--106},
}

@article{censor_dispersion_1980,
	title = {Dispersion equations in moving media},
	volume = {68},
	issn = {0018-9219},
	url = {http://ieeexplore.ieee.org/document/1455946/},
	doi = {10.1109/PROC.1980.11677},
	number = {4},
	urldate = {2024-02-14},
	journal = {Proc. IEEE},
	author = {Censor, D.},
	year = {1980},
	pages = {528--529},
}

@article{mccall_relativity_2007,
	title = {Relativity and mathematical tools: {Waves} in moving media},
	volume = {75},
	issn = {0002-9505, 1943-2909},
	shorttitle = {Relativity and mathematical tools},
	url = {https://pubs.aip.org/ajp/article/75/12/1134/899076/Relativity-and-mathematical-tools-Waves-in-moving},
	doi = {10.1119/1.2772281},
	number = {12},
	urldate = {2024-02-14},
	journal = {Am. J. Phys.},
	author = {McCall, Martin and Censor, Dan},
	month = dec,
	year = {2007},
	pages = {1134--1140},
}

@article{censor_relativistic_2010,
	title = {Relativistic invariance of dispersion‐relations and their associated wave‐operators and {Green}‐functions},
	volume = {90},
	issn = {0044-2267, 1521-4001},
	url = {https://onlinelibrary.wiley.com/doi/10.1002/zamm.200900298},
	doi = {10.1002/zamm.200900298},
	number = {3},
	urldate = {2024-02-14},
	journal = {ZAMM},
	author = {Censor, D.},
	month = mar,
	year = {2010},
	pages = {194--202},
}

@article{jones_rotary_1976,
	title = {Rotary ‘aether drag’},
	volume = {349},
	journal = {Proc. B. Soc. Lond. A.},
	author = {Jones, R. V.},
	year = {1976},
	pages = {423--439},
	doi = {10.1098/rspa.1976.0082}
}

@article{player_dragging_1976,
	title = {On the dragging of the plane of polarization of light propagating in a rotating medium},
	volume = {349},
	issn = {0080-4630},
	url = {https://royalsocietypublishing.org/doi/10.1098/rspa.1976.0083},
	doi = {10.1098/rspa.1976.0083},
	number = {1659},
	urldate = {2024-02-14},
	journal = {Proc. R. Soc. A},
	author = {Player, M. A.},
	month = jun,
	year = {1976},
	pages = {441--445},
}

@article{gueroult_determining_2019,
	title = {Determining the rotation direction in pulsars},
	volume = {10},
	issn = {2041-1723},
	url = {https://www.nature.com/articles/s41467-019-11243-4},
	doi = {10.1038/s41467-019-11243-4},
	number = {1},
	urldate = {2024-02-14},
	journal = {Nature Commun.},
	author = {Gueroult, Renaud and Shi, Yuan and Rax, Jean-Marcel and Fisch, Nathaniel J.},
	month = jul,
	year = {2019},
	pages = {3232},
}

@book{rax_physique_2005,
	address = {Paris},
	title = {Physique des plasmas: cours et applications},
	isbn = {978-2-10-007250-7},
	shorttitle = {Physique des plasmas},
	publisher = {Dunod},
	author = {Rax, J. M.},
	year = {2005}
}

@book{tracy_ray_2014,
	title = {Ray {Tracing} and {Beyond}: phase space methods in plasma wave theory},
	isbn = {978-0-521-76806-1},
	publisher = {Cambridge Univ. Press},
	author = {Tracy, E. R. and Brizard, A. J. and Richardson, A. S. and 
Kaufman, A. N.},
	year = {2014}
}

@article{ko_passage_1978,
	title = {On the passage of radiation through moving astrophysical plasmas},
	volume = {222},
	journal = {Astrophys. J.},
	author = {Ko, H. C. and Chuang, C. W.},
	year = {1978},
	pages = {1012--1019},
	doi = {10.1086/156219}
}

@article{ruiz_lagrangian_2015,
	title = {Lagrangian geometrical optics of nonadiabatic vector waves and spin particles},
	volume = {379},
	issn = {03759601},
	url = {https://linkinghub.elsevier.com/retrieve/pii/S0375960115006404},
	doi = {10.1016/j.physleta.2015.07.038},
	number = {38},
	urldate = {2024-05-22},
	journal = {Phys. Lett. A},
	author = {Ruiz, D.E. and Dodin, I.Y.},
	month = oct,
	year = {2015},
	pages = {2337--2350},
}

@article{dodin_quasioptical_2019,
	title = {Quasioptical modeling of wave beams with and without mode conversion. {I}. {Basic} theory},
	volume = {26},
	issn = {1070-664X, 1089-7674},
	url = {https://pubs.aip.org/pop/article/26/7/072110/1059086/Quasioptical-modeling-of-wave-beams-with-and},
	doi = {10.1063/1.5095076},
	number = {7},
	urldate = {2024-05-22},
	journal = {Phys. Plasma},
	author = {Dodin, I. Y. and Ruiz, D. E. and Yanagihara, K. and Zhou, Y. and Kubo, S.},
	month = jul,
	year = {2019},
	pages = {072110},
}

@article{fermi_rotation_1923,
	title = {On the rotation of the plane of polarization in a rotating medium},
	journal = {Rend. Mat. Acc. Lincei},
	author = {Fermi, E},
	volume = {32},
	pages = {115},
	year = {1923},
	note = {reprinted in \emph{Collected Papers of Enrico Fermi} (University of
Chicago Press, Chicago, 1962), Vol. 1}
}

@article{gedalin_waves_2001,
	title = {Waves in strongly magnetized relativistic plasmas: {Generally} covariant approach},
	volume = {64},
	copyright = {http://link.aps.org/licenses/aps-default-license},
	issn = {1063-651X, 1095-3787},
	shorttitle = {Waves in strongly magnetized relativistic plasmas},
	url = {https://link.aps.org/doi/10.1103/PhysRevE.64.027401},
	doi = {10.1103/PhysRevE.64.027401},
	number = {2},
	urldate = {2024-09-03},
	journal = {Phys. Rev. E},
	author = {Gedalin, M. and Melrose, D. B.},
	month = jul,
	year = {2001},
	pages = {027401},
}

@article{hebenstreit_calculation_1979,
	title = {Calculation of {Covariant} {Dispersion} {Equations} for {Moving} {Plasmas}},
	volume = {34},
	copyright = {http://creativecommons.org/licenses/by-nc-nd/3.0/},
	issn = {1865-7109, 0932-0784},
	url = {https://www.degruyter.com/document/doi/10.1515/zna-1979-0205/html},
	doi = {10.1515/zna-1979-0205},
	number = {2},
	urldate = {2024-09-05},
	journal = {Z. Naturforsch., A},
	author = {Hebenstreit, Helmut},
	month = feb,
	year = {1979},
	pages = {155--162},
}

@article{rytov_sur_1938,
	title = {On transition from wave to geometrical optics},
	volume = {18},
	number = {4-5},
	journal = {Dokl. Akad. Nauk SSSR},
	author = {Rytov, S. M.},
	year = {1938},
	pages = {263--266}
}

@article{berry_interpreting_1987,
	title = {Interpreting the anholonomy of coiled light},
	volume = {326},
	copyright = {http://www.springer.com/tdm},
	issn = {0028-0836, 1476-4687},
	url = {https://www.nature.com/articles/326277a0},
	doi = {10.1038/326277a0},
	number = {6110},
	urldate = {2024-10-07},
	journal = {Nature},
	author = {Berry, M. V.},
	month = mar,
	year = {1987},
	pages = {277--278},
}

@article{liberman_spin-orbit_1992,
	title = {Spin-orbit interaction of a photon in an inhomogeneous medium},
	volume = {46},
	copyright = {http://link.aps.org/licenses/aps-default-license},
	issn = {1050-2947, 1094-1622},
	url = {https://link.aps.org/doi/10.1103/PhysRevA.46.5199},
	doi = {10.1103/PhysRevA.46.5199},
	number = {8},
	urldate = {2024-10-07},
	journal = {Phys. Rev. A},
	author = {Liberman, V. S. and Zel’dovich, B. Ya.},
	month = oct,
	year = {1992},
	pages = {5199--5207},
}

@article{berry_quantal_1984,
	title = {Quantal {Phase} {Factors} {Accompanying} {Adiabatic} {Changes}},
	author = {Berry, M V},
	year = {1984},
	journal = {Proc. R. Soc. A},
	volume = {392},
	number = {1802},
	pages = {45--57},
	doi = {10.1098/rspa.1984.0023}
}

@article{samuel_general_1988,
	title = {General setting for {Berry}'s phase},
	volume = {60},
	number = {23},
	journal = {Phys. Rev. Lett.},
	author = {Samuel, Joseph and Bhandari, Rajendra},
	year = {1988},
	pages = {2339},
	doi = {10.1103/PhysRevLett.60.2339}
}

@article{berry_budden_1990,
	title = {Budden \& {Smith}'s `{Additional} {Memory}' and the {Geometric} {Phase}},
	volume = {431},
	number = {1883},
	journal = {Proc. R. Soc. A},
	author = {Berry, M V},
	year = {1990},
	pages = {531--537},
	doi = {10.1098/rspa.1990.0149}
}

@article{cisowski_colloquium_2022,
	title = {\textit{{Colloquium}} : {Geometric} phases of light: {Insights} from fiber bundle theory},
	volume = {94},
	issn = {0034-6861, 1539-0756},
	shorttitle = {\textit{{Colloquium}}},
	url = {https://link.aps.org/doi/10.1103/RevModPhys.94.031001},
	doi = {10.1103/RevModPhys.94.031001},
	number = {3},
	urldate = {2025-05-21},
	journal = {Rev. Mod. Phys.},
	author = {Cisowski, C. and Götte, J. B. and Franke-Arnold, S.},
	month = jul,
	year = {2022},
	pages = {031001},
}

@article{braud_spin-orbit_2025,
	title = {Spin-orbit interactions induced by light drag in moving media},
	doi = {10.1103/lsgb-ylzf},
	journal = {Phys. Rev. A},
	author = {Braud, Aymeric and Gueroult, Renaud},
	year = {2025},
	volume = {112},
	pages = {043505}
}

@article{vinitskii_topological_1990,
	title = {Topological phases in quantum mechanics and polarization optics},
	author = {Vinitskii, S I and Derbov, V L and Dubovik, V N and Markovski, B L and Stepanovskii, Yu P},
	year = {1990},
	journal = {Usp. Fiz. Nauk.},
	volume = {160},
	pages = {1--49},
	doi = {10.1070/PU1990v033n06ABEH002598}
}

@book{born_principles_2019,
	title = {Principles of optics},
	isbn = {978-1-108-47743-7 978-1-108-76991-4},
	publisher = {Cambridge Univ. Press},
	author = {Born, Max and Wolf, Emil and Knight, Peter},
	collaborator = {Bhatia, A. B.},
	year = {2019},
	doi = {10.1017/9781108769914}
}

@book{anile_relativistic_2005,
	series = {Cambridge monographs on mathematical physics},
	title = {Relativistic fluids and magneto-fluids: with applications in astrophysics and plasma physics},
	isbn = {978-0-521-30406-1 978-0-521-01812-8},
	publisher = {Cambridge Univ. Press},
	author = {Anile, Angelo Marcello},
	year = {2005},
}

@article{tam_covariant_1968,
	title = {A covariant treatment of relativistic plasma oscillations},
	volume = {46},
	copyright = {http://www.nrcresearchpress.com/page/about/CorporateTextAndDataMining},
	issn = {0008-4204, 1208-6045},
	url = {https://cdnsciencepub.com/doi/10.1139/p68-511},
	doi = {10.1139/p68-511},
	number = {16},
	urldate = {2025-11-25},
	journal = {Can. J. Phys.},
	author = {Tam, Kwok-Kee},
	month = aug,
	year = {1968},
	pages = {1763--1767},
}

@Article{Berard2006,
  author    = {Bérard, Alain and Mohrbach, Hervé},
  journal   = {Phys. Lett. A},
  title     = {Spin Hall effect and Berry phase of spinning particles},
  year      = {2006},
  issn      = {0375-9601},
  month     = mar,
  number    = {3},
  pages     = {190--195},
  volume    = {352},
  doi       = {10.1016/j.physleta.2005.11.071},
  publisher = {Elsevier BV},
}

@Article{Cardano2015,
  author    = {Cardano, Filippo and Marrucci, Lorenzo},
  journal   = {Nat. Photonics},
  title     = {Spin–orbit photonics},
  year      = {2015},
  issn      = {1749-4893},
  month     = nov,
  number    = {12},
  pages     = {776--778},
  volume    = {9},
  doi       = {10.1038/nphoton.2015.232},
  publisher = {Springer Science and Business Media LLC},
}

@Article{Mathur1991,
  author    = {Mathur, Harsh},
  journal   = {Phys. Rev. Lett.},
  title     = {Thomas precession, spin-orbit interaction, and Berry’s phase},
  year      = {1991},
  issn      = {0031-9007},
  month     = dec,
  number    = {24},
  pages     = {3325--3327},
  volume    = {67},
  doi       = {10.1103/physrevlett.67.3325},
  publisher = {American Physical Society (APS)},
}

@Article{Sheng2023,
  author    = {Sheng, Lijuan and Chen, Yu and Yuan, Shuaijie and Liu, Xuquan and Zhang, Zhiyou and Jing, Hui and Kuang, Le-Man and Zhou, Xinxing},
  journal   = {Prog. Quantum Electron.},
  title     = {Photonic spin Hall effect: Physics, manipulations, and applications},
  year      = {2023},
  issn      = {0079-6727},
  month     = nov,
  pages     = {100484},
  volume    = {91–92},
  doi       = {10.1016/j.pquantelec.2023.100484},
  publisher = {Elsevier BV},
}

@Article{Ling2017,
  author    = {Ling, Xiaohui and Zhou, Xinxing and Huang, Kun and Liu, Yachao and Qiu, Cheng-Wei and Luo, Hailu and Wen, Shuangchun},
  journal   = {Rep. Progr. Phys.},
  title     = {Recent advances in the spin Hall effect of light},
  year      = {2017},
  issn      = {1361-6633},
  month     = mar,
  number    = {6},
  pages     = {066401},
  volume    = {80},
  doi       = {10.1088/1361-6633/aa5397},
  publisher = {IOP Publishing},
}

@Article{Allen1992,
  author    = {Allen, L. and Beijersbergen, M. W. and Spreeuw, R. J. C. and Woerdman, J. P.},
  journal   = {Phys. Rev. A},
  title     = {Orbital angular momentum of light and the transformation of Laguerre-Gaussian laser modes},
  year      = {1992},
  issn      = {1094-1622},
  month     = jun,
  number    = {11},
  pages     = {8185--8189},
  volume    = {45},
  doi       = {10.1103/physreva.45.8185},
  publisher = {American Physical Society (APS)},
}

@Article{Poynting1909,
  author  = {Poynting, John Henry },
  journal = {Proc. R. Soc. Lond. A.},
  title   = {The wave motion of a revolving shaft, and a suggestion as to the angular momentum in a beam of circularly polarised light},
  year    = {1909},
  number  = {557},
  pages   = {560-567},
  volume  = {82},
  doi     = {10.1098/rspa.1909.0060},
}

@Article{Vladimirskiy1941,
  author  = {Vladimirskiy, V. V.},
  journal = {Dokl. Akad. Nauk SSSR},
  title   = {The rotation of a polarization plane for curved light ray},
  year    = {1941},
  volume = {21},
  pages = {222}
}

@Article{Beth1936,
  author    = {Beth, Richard A.},
  journal   = {Phys. Rev.},
  title     = {Mechanical Detection and Measurement of the Angular Momentum of Light},
  year      = {1936},
  issn      = {0031-899X},
  month     = jul,
  number    = {2},
  pages     = {115--125},
  volume    = {50},
  doi       = {10.1103/physrev.50.115},
  publisher = {American Physical Society (APS)},
}

@Article{Allen1996,
  author    = {Allen, L. and Lembessis, V. E. and Babiker, M.},
  journal   = {Phys. Rev. A},
  title     = {Spin-orbit coupling in free-space Laguerre-Gaussian light beams},
  year      = {1996},
  issn      = {1094-1622},
  month     = may,
  number    = {5},
  pages     = {R2937–R2939},
  volume    = {53},
  doi       = {10.1103/physreva.53.r2937},
  publisher = {American Physical Society (APS)},
}

@Article{Zhang2024,
  author    = {Zhang, Ze-Lin and Liu, Ruo-Yu},
  journal   = {Astrophys. J.},
  title     = {The Twisting of Radio Waves in a Randomly Inhomogeneous Plasma},
  year      = {2024},
  issn      = {1538-4357},
  month     = nov,
  number    = {2},
  pages     = {260},
  volume    = {975},
  doi       = {10.3847/1538-4357/ad7d0a},
  publisher = {American Astronomical Society},
}

@Article{Tomita1986,
  author    = {Tomita, Akira and Chiao, Raymond Y.},
  journal   = {Phys. Rev. Lett.},
  title     = {Observation of Berry’s Topological Phase by Use of an Optical Fiber},
  year      = {1986},
  issn      = {0031-9007},
  month     = aug,
  number    = {8},
  pages     = {937--940},
  volume    = {57},
  doi       = {10.1103/physrevlett.57.937},
  publisher = {American Physical Society (APS)},
}

@Article{Ross1984,
  author    = {Ross, J. N.},
  journal   = {Opt. Quantum Electron.},
  title     = {The rotation of the polarization in low birefringence monomode optical fibres due to geometric effects},
  year      = {1984},
  issn      = {1572-817X},
  month     = sep,
  number    = {5},
  pages     = {455--461},
  volume    = {16},
  doi       = {10.1007/bf00619638},
  publisher = {Springer Science and Business Media LLC},
}

@Article{Minkowski1908,
  author  = {Minkowski, H.},
  journal = {Nachr. Ges. Wiss. Gottingen, Math.-Phys. Kl.},
  title   = {Die Grundgleichungen für die elektromagnetischen Vorgänge in bewegten Körpern},
  year    = {1908},
  pages   = {53-111},
  volume  = {1908},
  url     = {http://eudml.org/doc/58707},
}

@Book{Kong2000,
  author    = {Kong, J. A.},
  publisher = {EMW, Cambridge, MA},
  title     = {Electromagnetic wave theory},
  year      = {2000},
}

@Article{Braud2025,
  author  = {Braud, A. and Langlois, J. and Gueroult, R.},
  journal = {C.R. Phys.},
  title   = {Geometrical optics methods for moving anisotropic media: a tool for magnetized plasmas},
  year    = {2025},
  pages   = {7},
  volume  = {26},
  doi     = {10.5802/crphys.218},
}

@Article{Arnaud1976,
  author    = {Arnaud, J. A.},
  journal   = {Nature},
  title     = {Dispersion and the transverse aether drag},
  year      = {1976},
  issn      = {1476-4687},
  month     = jun,
  number    = {5560},
  pages     = {481--482},
  volume    = {261},
  doi       = {10.1038/261481a0},
  publisher = {Springer Science and Business Media LLC},
}

@Article{Langlois2024,
  author    = {Langlois, Julien and Gueroult, Renaud},
  journal   = {Proc. R. Soc. A.},
  title     = {Manifestations of inertia on light dragging revealed in plasmas},
  year      = {2024},
  issn      = {1471-2946},
  month     = nov,
  number    = {2301},
  pages     = {20240300},
  volume    = {480},
  doi       = {10.1098/rspa.2024.0300},
  publisher = {The Royal Society},
}

@Article{Leclerc2025,
  author    = {Leclerc, Armand and Laibe, Guillaume},
  journal   = {Astrophys. J.},
  title     = {The Importance of Berry Phase in Solar Acoustic Modes},
  year      = {2025},
  issn      = {2041-8213},
  month     = apr,
  number    = {1},
  pages     = {L17},
  volume    = {983},
  doi       = {10.3847/2041-8213/adc457},
  publisher = {American Astronomical Society},
}

@Article{Torabi2012,
  author    = {Torabi, R. and Mehrafarin, M.},
  journal   = {JETP Lett.},
  title     = {Berry effect in unmagnetized inhomogeneous cold plasmas},
  year      = {2012},
  issn      = {1090-6487},
  month     = may,
  number    = {6},
  pages     = {277--281},
  volume    = {95},
  doi       = {10.1134/s0021364012060112},
  publisher = {Pleiades Publishing Ltd},
}

@Article{Oancea2024,
  author    = {Oancea, Marius A and Stiskalek, Richard and Zumalacárregui, Miguel},
  journal   = {MNRAS: Letters},
  title     = {Probing general relativistic spin–orbit coupling with gravitational waves from hierarchical triple systems},
  year      = {2024},
  issn      = {1745-3933},
  month     = sep,
  number    = {1},
  pages     = {L1--L6},
  volume    = {535},
  doi       = {10.1093/mnrasl/slae084},
  publisher = {Oxford University Press (OUP)},
}

@Article{Gueroult2025,
  author    = {Gueroult, R. and Tripathi, S. K. and Han, J. and Pribyl, P. and Rax, J.-M. and Fisch, N. J.},
  journal   = {Phys. Rev. Lett.},
  title     = {Image rotation in plasmas},
  year      = {2025},
  issn      = {1079-7114},
  month     = may,
  pages     = {245101},
  volume    = {134},
  doi       = {10.1103/swrn-w3yf},
  publisher = {American Physical Society (APS)},
}

@Article{Berry2001,
  author    = {Berry, M.V and Dennis, M.R},
  journal   = {Proc. R. Soc. Lond. A},
  title     = {Polarization singularities in isotropic random vector waves},
  year      = {2001},
  issn      = {1471-2946},
  month     = jan,
  number    = {2005},
  pages     = {141--155},
  volume    = {457},
  doi       = {10.1098/rspa.2000.0660},
  publisher = {The Royal Society},
}

@Article{Bliokh2014,
  author    = {Bliokh, Konstantin Y. and Bekshaev, Aleksandr Y. and Nori, Franco},
  journal   = {Nat. Commun.},
  title     = {Extraordinary momentum and spin in evanescent waves},
  year      = {2014},
  issn      = {2041-1723},
  pages     = {3300},
  month     = mar,
  number    = {1},
  volume    = {5},
  doi       = {10.1038/ncomms4300},
  publisher = {Springer Science and Business Media LLC},
}

@Article{Neugebauer2018,
  author    = {Neugebauer, Martin and Eismann, Jörg S. and Bauer, Thomas and Banzer, Peter},
  journal   = {Phys. Rev. X},
  title     = {Magnetic and Electric Transverse Spin Density of Spatially Confined Light},
  year      = {2018},
  issn      = {2160-3308},
  month     = may,
  number    = {2},
  pages     = {021042},
  volume    = {8},
  doi       = {10.1103/physrevx.8.021042},
  publisher = {American Physical Society (APS)},
}

@Article{Aiello2015,
  author    = {Aiello, Andrea and Banzer, Peter and Neugebauer, Martin and Leuchs, Gerd},
  journal   = {Nat. Photonics},
  title     = {From transverse angular momentum to photonic wheels},
  year      = {2015},
  issn      = {1749-4893},
  month     = nov,
  number    = {12},
  pages     = {789--795},
  volume    = {9},
  doi       = {10.1038/nphoton.2015.203},
  publisher = {Springer Science and Business Media LLC},
}

@Article{Peng2019,
  author    = {Peng, Liang and Duan, Lingfu and Wang, Kewen and Gao, Fei and Zhang, Li and Wang, Gaofeng and Yang, Yihao and Chen, Hongsheng and Zhang, Shuang},
  journal   = {Nat. Photonics},
  title     = {Transverse photon spin of bulk electromagnetic waves in bianisotropic media},
  year      = {2019},
  issn      = {1749-4893},
  month     = oct,
  number    = {12},
  pages     = {878--882},
  volume    = {13},
  doi       = {10.1038/s41566-019-0521-4},
  publisher = {Springer Science and Business Media LLC},
}

@Book{Booker1984,
  author    = {Booker, H. G.},
  publisher = {Springer Science \& Business Media},
  title     = {Cold Plasma Waves},
  year      = {1984},
}

@Article{Gong2021,
  author    = {Gong, Su-Hyun and Park, Q-Han},
  journal   = {Opt. Express},
  title     = {Gyroelectric guided modes with transverse optical spin},
  year      = {2021},
  issn      = {1094-4087},
  month     = mar,
  number    = {7},
  pages     = {10631},
  volume    = {29},
  doi       = {10.1364/oe.421548},
  publisher = {Optica Publishing Group},
}

@Article{Neugebauer2015,
  author    = {Neugebauer, Martin and Bauer, Thomas and Aiello, Andrea and Banzer, Peter},
  journal   = {Phys. Rev. Lett.},
  title     = {Measuring the Transverse Spin Density of Light},
  year      = {2015},
  issn      = {1079-7114},
  month     = feb,
  number    = {6},
  pages     = {063901},
  volume    = {114},
  doi       = {10.1103/physrevlett.114.063901},
  publisher = {American Physical Society (APS)},
}

@Article{Chen2017,
  author    = {Chen, Jian and Wan, Chenhao and Kong, Ling Jiang and Zhan, Qiwen},
  journal   = {Opt. Express},
  title     = {Tightly focused optical field with controllable photonic spin orientation},
  year      = {2017},
  issn      = {1094-4087},
  month     = aug,
  number    = {16},
  pages     = {19517},
  volume    = {25},
  doi       = {10.1364/oe.25.019517},
  publisher = {Optica Publishing Group},
}

@Article{Bliokh2015a,
  author    = {Bliokh, Konstantin Y. and Nori, Franco},
  journal   = {Phys. Rep.},
  title     = {Transverse and longitudinal angular momenta of light},
  year      = {2015},
  issn      = {0370-1573},
  month     = aug,
  pages     = {1--38},
  volume    = {592},
  doi       = {10.1016/j.physrep.2015.06.003},
  publisher = {Elsevier BV},
}

@Article{Fu2024,
  author    = {Fu, Tong and Lin, Jiaxin and Xu, Yuhao and Jia, Junji and Wang, Yonglong and Zhang, Shunping and Xu, Hongxing},
  journal   = {Nano Lett.},
  title     = {Transverse Spin–Orbit Interaction of Light},
  year      = {2024},
  issn      = {1530-6992},
  month     = aug,
  number    = {35},
  pages     = {10783--10789},
  volume    = {24},
  doi       = {10.1021/acs.nanolett.4c01931},
  publisher = {American Chemical Society (ACS)},
}

@Article{Bekshaev2015,
  author    = {Bekshaev, Aleksandr Y. and Bliokh, Konstantin Y. and Nori, Franco},
  journal   = {Phys. Rev. X},
  title     = {Transverse Spin and Momentum in Two-Wave Interference},
  year      = {2015},
  issn      = {2160-3308},
  month     = mar,
  number    = {1},
  pages     = {011039},
  volume    = {5},
  doi       = {10.1103/physrevx.5.011039},
  publisher = {American Physical Society (APS)},
}

@Article{CanaguierDurand2014,
  author    = {Canaguier-Durand, Antoine and Genet, Cyriaque},
  journal   = {Phys. Rev. A},
  title     = {Transverse spinning of a sphere in a plasmonic field},
  year      = {2014},
  issn      = {1094-1622},
  month     = mar,
  number    = {3},
  pages     = {033841},
  volume    = {89},
  doi       = {10.1103/physreva.89.033841},
  publisher = {American Physical Society (APS)},
}

@Article{Bliokh2012,
  author    = {Bliokh, Konstantin Y. and Nori, Franco},
  journal   = {Phys. Rev. A},
  title     = {Transverse spin of a surface polariton},
  year      = {2012},
  issn      = {1094-1622},
  month     = jun,
  number    = {6},
  pages     = {061801},
  volume    = {85},
  doi       = {10.1103/physreva.85.061801},
  publisher = {American Physical Society (APS)},
}

@Article{Bliokh2019,
  author    = {Bliokh, Konstantin Y and Alonso, Miguel A and Dennis, Mark R},
  journal   = {Rep. Progr. Phys.},
  title     = {Geometric phases in 2D and 3D polarized fields: geometrical, dynamical, and topological aspects},
  year      = {2019},
  issn      = {1361-6633},
  month     = oct,
  number    = {12},
  pages     = {122401},
  volume    = {82},
  doi       = {10.1088/1361-6633/ab4415},
  publisher = {IOP Publishing},
}

@Article{Liu2016,
  author    = {Liu, Yachao and Ke, Yougang and Luo, Hailu and Wen, Shuangchun},
  journal   = {Nanophotonics},
  title     = {Photonic spin Hall effect in metasurfaces: a brief review},
  year      = {2016},
  issn      = {2192-8606},
  month     = jul,
  number    = {1},
  pages     = {51--70},
  volume    = {6},
  doi       = {10.1515/nanoph-2015-0155},
  publisher = {Walter de Gruyter GmbH},
}

\end{document}